\documentclass[useAMS,usenatbib,usegraphicx]{mn2e}

\def \OIII {[O\,{\sc iii}]\ $\lambda$5007\,\AA}
\def \NII {[N\,{\sc ii}]\ $\lambda$6584\,\AA}
\def \ha {H$\alpha$}

\def \vhel{\ifmmode{~V_{{\rm HEL}}}\else{~$V_{{\rm HEL}}$}\fi}
\def\msun{\ifmmode{{\rm\ M}_\odot}\else{${\rm\ M}_\odot$}\fi}

\def\myr{\ifmmode{{\rm\ M}_\odot{\rm\ yr}^{-1}}
         \else{${\rm\ M}_\odot$ yr$^{-1}$}\fi}
\def\kms{\ifmmode{~{\rm km\,s}^{-1}}\else{~km s$^{-1}$}\fi}

\title[The kinematics and shaping of PN HaTr~4]{A study of the kinematics and binary-induced shaping of the planetary nebula HaTr~4\thanks{Based on observations made with European Southern Observatory Telescopes at the La Silla Paranal Observatory, under program IDs 081.D-0857 and 055.D-0550}}
\author[A.A. Tyndall et al.]{A.A. Tyndall,$^{1}$$^,$$^{2}$$^,$$^{3}$\thanks{E-mail:
amy.tyndall@postgrad.manchester.ac.uk} D. Jones,$^{3}$ M. Lloyd,$^{1}$ T.J. O'Brien,$^{1}$and  D. Pollacco$^{4}$ \\
\\
$^{1}$Jodrell Bank Centre for Astrophysics, School of Physics and Astronomy, University of Manchester, M13 9PL, UK\\
$^{2}$Isaac Newton Group of Telescopes, Apartado de Correos 368, E-38700 Santa Cruz de La Palma, Spain\\
$^{3}$European Southern Observatory, Alonso de C\'ordova 3107, Casilla 19001, Santiago, Chile\\ 
$^{4}$Astrophysics Research Centre, Queen's University Belfast, BT7 1NN, UK}

\begin{document}

\date{Accepted xxxx xxxxxxxx xx. Received xxxx xxxxxxxx xx; in original form xxxx xxxxxxxx xx}

\pagerange{\pageref{firstpage}--\pageref{lastpage}} \pubyear{2010}

\maketitle

\label{firstpage}

\begin{abstract}

We present the first detailed spatio-kinematical analysis and modelling of the planetary nebula HaTr~4, one of few known to contain a post-common-envelope central star system. Common envelope evolution is believed to play an important role in the shaping of planetary nebulae, but the exact nature of this role is yet to be understood. High spatial- and spectral- resolution spectroscopy of the \OIII{} nebular line obtained with VLT-UVES are presented alongside deep narrowband \ha+\NII{} imagery obtained using EMMI-NTT, and together the two are used to derive the three-dimensional morphology of HaTr~4. The nebula is found to display an extended ovoid morphology with an enhanced equatorial region consistent with a toroidal waist - a feature believed to be typical amongst planetary nebulae with post-common-envelope central stars. The nebular symmetry axis is found to lie perpendicular to the orbital plane of the central binary, concordant with the idea that the formation and evolution of HaTr~4 has been strongly influenced by its central binary.
%HaTr~4 is one of very few nebulae to have this predicted alignment between nebular and binary plane observationally proven.

\end{abstract}


\section{Introduction}

The current understanding of planetary nebulae (PNe) is that they are formed when low-to-intermediate mass stars of $0.8M_\odot$ $<$ M $<$ $8M_\odot$ start to transition from an asymptotic-giant-branch (AGB) star to a white dwarf. \citet{kwok78} first put forward the `Interacting Stellar Winds'\ (ISW) theory of nebular formation, whereby a PN is formed by a `snow-plow'\ process. Here, a slow, dense `superwind', blown while the central star is in the AGB phase, is swept up by a fast, tenuous wind emanating from the emerging white dwarf creating a compressed, thin, dense shell. This shell is then ionised by the hot central star to produce the visible PN. Balick \& Frank \citeyearpar{balick02} state that in general, the mass loss in these circumstances is inherently isotropic which should lead to the formation of a spherical nebula around the progenitor star. However, many PNe instead tend to show more complex, axisymmetric (bipolar) morphologies which contradict this theory. Within the framework of the ISW theory, any deviations from sphericity - particularly bipolar shapes - require an aspherical mass distribution in either wind phase \citep{kahn85}. This has become known as the `Generalised Interacting Stellar Winds'\ model (GISW, \citealp{balick02}). However, the source of this anisotropy is currently a matter of some controversy.

A strong argument is starting to be built up in favour of bipolar nebulae being shaped by a central binary star system (\citealp{demarco09a}, \citealp{miszalski09}, \citealp{jones11}). \citet{paczynski76} first identified a binary pathway to PN formation via a common-envelope phase, with the first such nebula discovered being Abell 63 \citep{bond78}. Here, the density contrast required by the GISW model to form aspherical morphologies arises from the ejection of the common envelope (CE) \citep{nordhaus06}. Frictional forces between the binary and CE result in the transfer of angular momentum, causing the CE to be preferentially ejected in the stellar orbital plane -- the symmetry axis of the PN is then expected to be perpendicular to the orbital plane of the binary. Therefore, observing this predicted alignment in a given PN is strongly indicative that the nebular morphology has been influenced by the binary central star.

Such is the strength of support for the role of binarity in the formation and evolution of PNe, that the working group PLAN-B (PLAnetary Nebula Binaries\footnote{http://www.wiyn.org/planb/}) has been formed in order to coordinate the research effort in this field.

To date, only five PNe have been observationally shown to possess the predicted alignment between nebula and binary plane -- Abell~63 \citep{mitchell07}, Abell~41 \citep{jones10a}, Abell~65, \citep{huckvale11,shimansky09}, NGC~6337 \citep{hillwig10,garcia09} and Sp~1 \citep{jones11a}.

PN HaTr~4 (PN G335.2-03.6, $\alpha = 16^{h}$ 45$^{m}$ 00.2$^{s}$, $\delta = -51\degr$ 12\arcmin{}  22.0\arcsec{}, J2000) is known to contain a photometric binary central star with a period of 1.71 days \citep{bond90}. Further, more detailed investigation by Hillwig et al. (in prep) and \citet{bodman12} revised the period to 1.74 days. Additionally, they find no evidence for eclipses in the lightcurve indicating that the inclination of the binary orbital plane with respect to the line of sight must be less than 85\degr{}, with the best fit to the data being from a model at 75\degr{} \citep{bodman12}.

Based on previous imagery - the discovery image of \citet{hartl85} and subsequent narrowband images of \citet{bond90} - HaTr~4 has the appearance of a ``classical butterfly'' bipolar nebula lying in the plane of the sky with twin lobes emanating from the central star in an East-West direction, as well as an indication of possessing fainter, elongated, open-ended lobes extending in a North-South direction (only revealed in later, deeper imagery - see section \ref{sec:obs}, figure \ref{fig:allhatr4}). However, imagery alone is insufficient to precisely and unambiguously determine the nebular structure and orientation, as PNe are very prone to inclination-dependent projection effects \citep{kwok10,frank93}. Only by using high-resolution, spatially resolved spectroscopy is it possible to fully ascertain the intrinsic three-dimensional structure of a given nebula in order to assess whether the binary has played a role in the shaping of the nebula. 

In this paper, we present long-slit spectroscopy and deep, narrow-band imagery of HaTr~4, from which a spatio-kinematical model is derived, in order to investigate the relationship between HaTr~4 and its central binary star. 
 
\section{Observations}
\label{sec:obs}

%\begin{figure*}
%\centering
%\includegraphics[scale=0.7]{model.eps}
%\caption{Low}
%\label{fig:meh}
%\end{figure*}
 
%1800s, NTT-EMMI, Bell & Pollacco, mention how hi-res image of PN from Don's NTT run was acquired

\begin{figure*}
\centering
\includegraphics[scale=1.3]{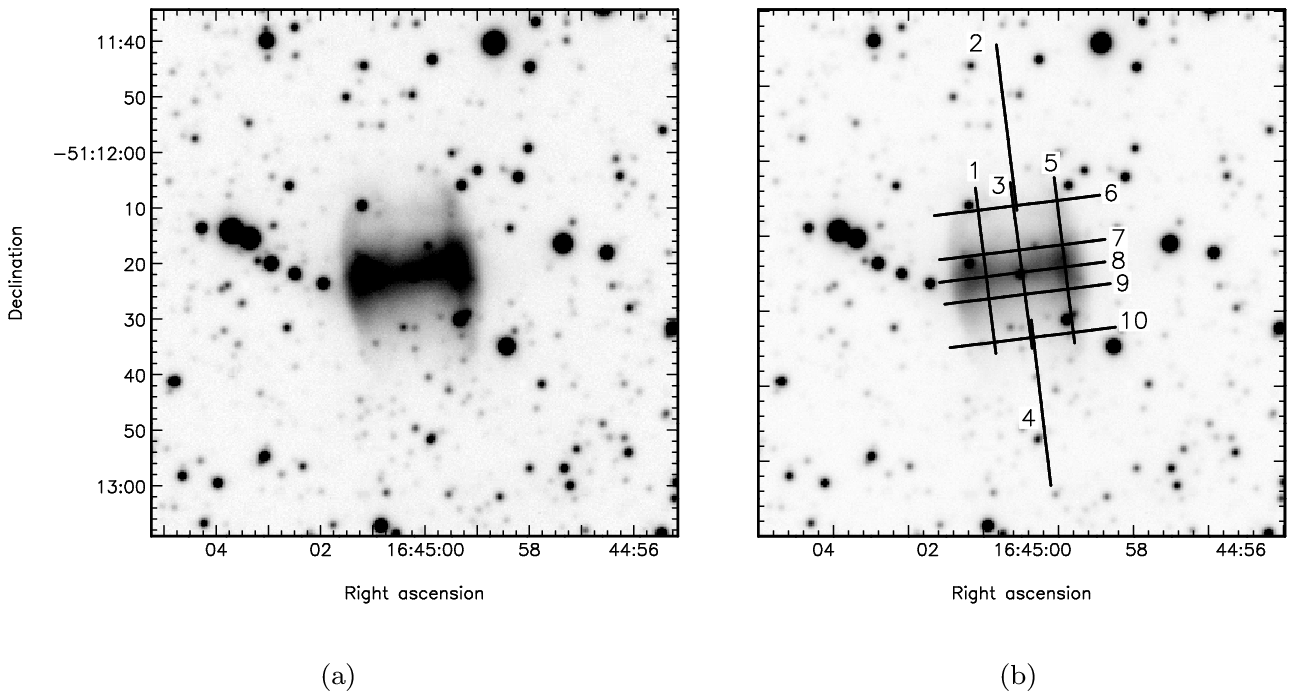}
\caption{(a) High contrast image of PN HaTr~4 taken on the NTT-EMMI spectrograph in \ha{}+[N~\textsc{ii}]. (b) The same NTT image presented at a lower contrast showing all slit positions taken on the VLT in [O~\textsc{iii}]. Slit width is 0.6\arcsec{}. The central star is visible at 16:45:00.2 -51:12:22.0}
\label{fig:allhatr4}
\end{figure*}

The deep \ha{}+\NII{} image shown in figure \ref{fig:allhatr4} was acquired on 1995 April 22, using the ESO (European Southern Observatory) Multi-Mode Instrument (EMMI; \citealt{dekker86}) on the 3.6-m ESO New Technology Telescope (NTT) with an exposure time of 1800~s and seeing of 1.1\arcsec{}. The image clearly shows the ``bow-tie''-shaped central nebula as seen in the \OIII{} imagery of \citet{bond90}, but also faint extensions to the North and South. These extensions are more consistent with a symmetry axis orientated roughly North-South (as opposed to the previously inferred East-West axis), indicative that the ``bow-tie'' may, in fact, form the waist of the nebula.

On 2008 July 10 and 11, data were acquired of HaTr~4 using grating \#3 on the visual-to-red arm of the Ultraviolet and Visual Echelle Spectrograph (UVES; \citealp{dekker00}) on the Keuyen Unit Telescope (UT2) of the Very Large Telescope (VLT). UVES was operated in its 30\arcsec{} longslit mode with a 0.6\arcsec{} slitwidth (R $\sim70,000$) using a narrowband filter to isolate the [O~\textsc{iii}] emission line profile and prevent contamination from overlapping orders. 1200~s exposures were taken at ten different slit positions (shown in figure \ref{fig:allhatr4}(b)). Slits 1--5 were taken at a position angle (P.A.) of 7$^{\circ}$, and slits 6--10 were taken at a P.A. of 97$^{\circ}$ in order to give full nebular coverage in both an East-West and North-South direction. The spatial scale of the observations was 0.17\arcsec{} per pixel, and the seeing was between 0.8\arcsec{} and 1.0\arcsec{} for all observations. 

\begin{figure*}
\centering
\includegraphics[]{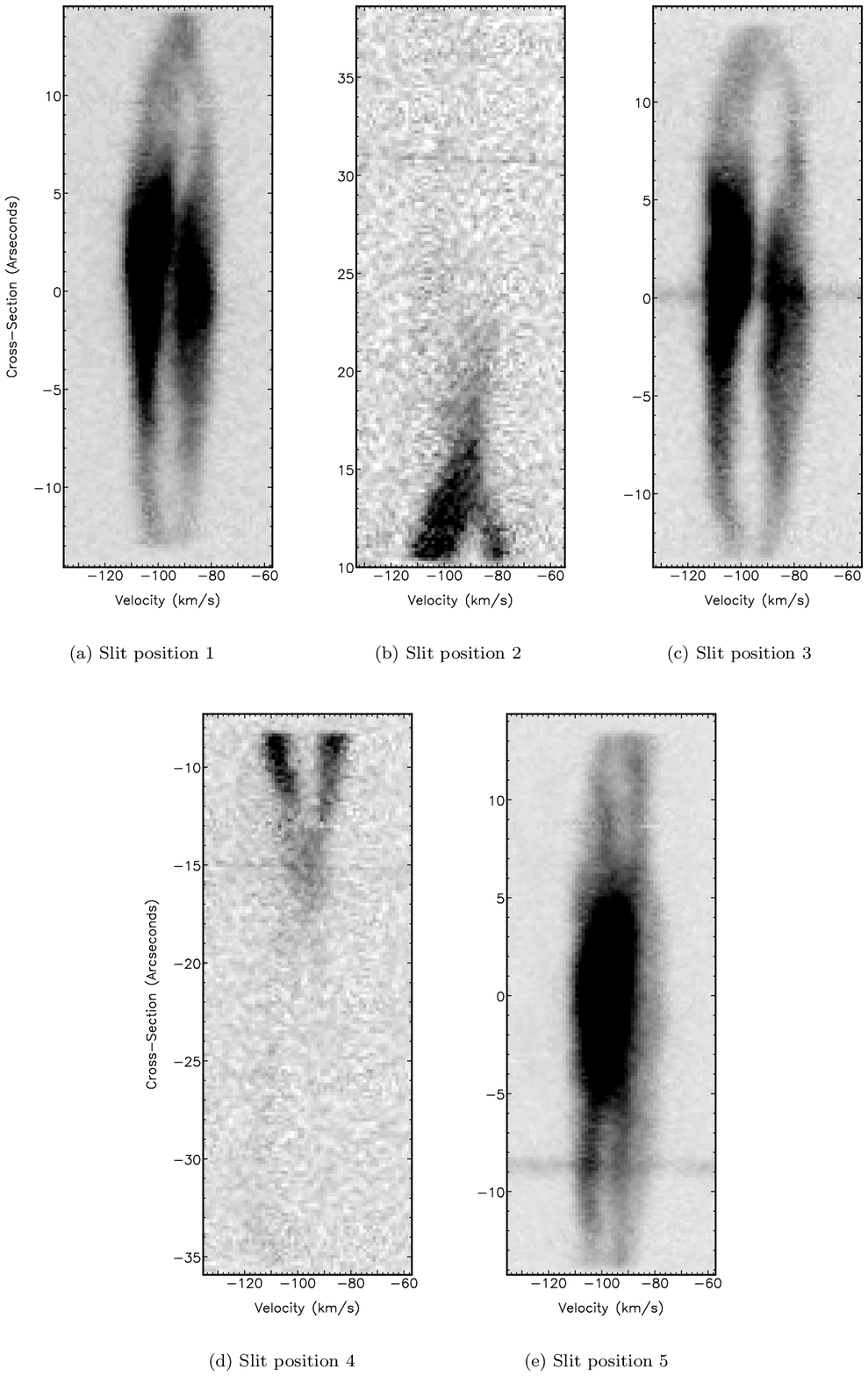}
\caption{PV arrays showing reduced [O~\textsc{iii}] spectra from the VLT slit positions 1-5. North is up, South is down. The velocity axis on all plots is heliocentric velocity, $v_{hel}$.  The display scales of each slit have been set individually to highlight the spatio-kinematic features referred to in the text. Cross-section 0\arcsec{} defines where the central star is found, the continuum of which can be seen in slit 3. The continuum of a field star is visible in slit 5 at cross-section -8\arcsec{}.}
\label{fig:1to5}
\end{figure*}

\begin{figure*}
\centering
\includegraphics[]{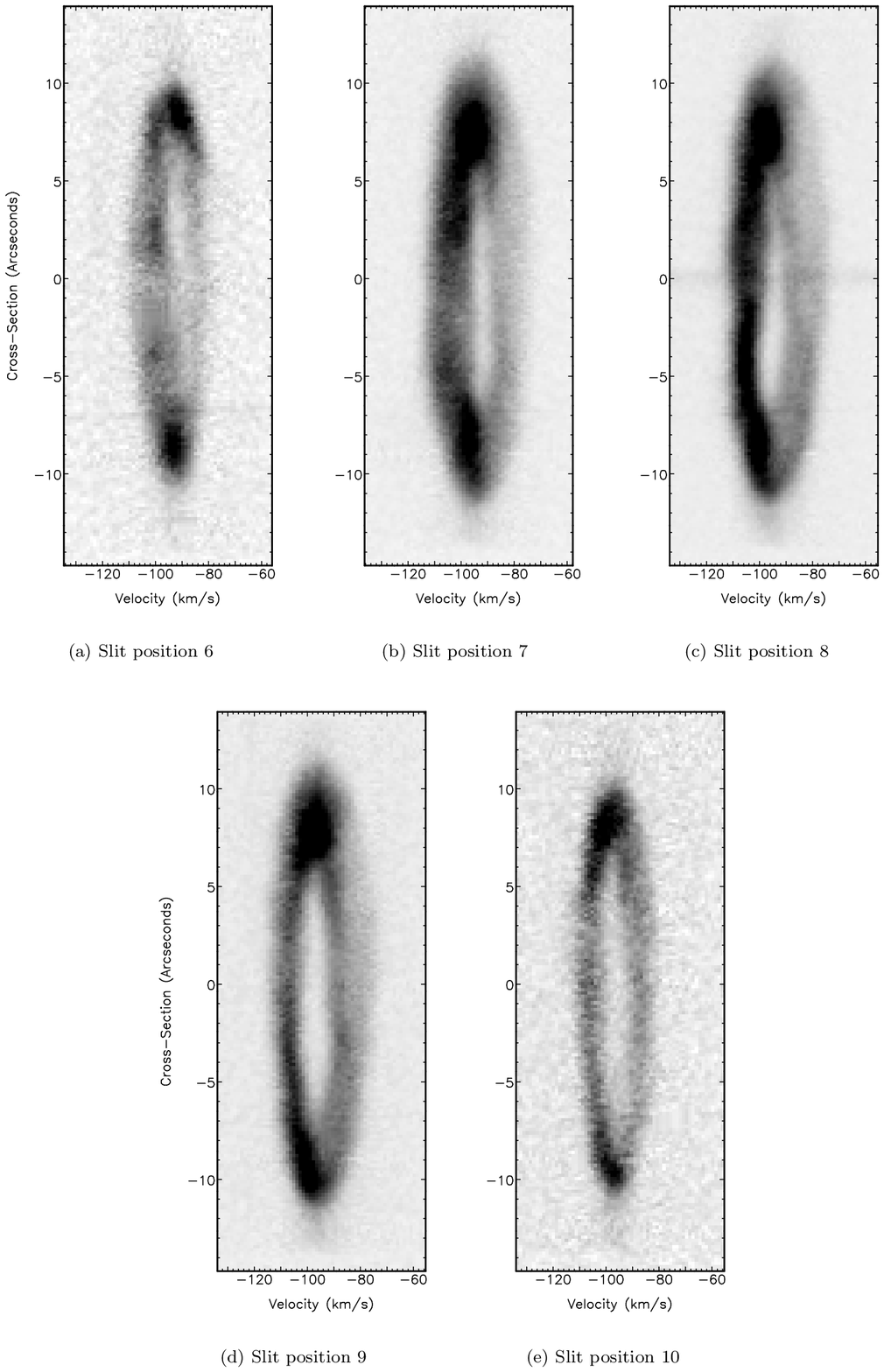}
\caption{PV arrays showing reduced [O~\textsc{iii}] spectra from the VLT slit positions 6-10. West is up, East is down. The velocity axis on all plots is heliocentric velocity, $v_{hel}$. The display scales of each slit have been set individually to highlight the spatio-kinematic features referred to in the text. Cross-section 0\arcsec{} defines where the central star is found.}
\label{fig:6to10}
\end{figure*}

Reduction of the spectra was carried out using standard routines within the \textsc{starlink} software package. The spectra were cleaned of cosmic rays and debiased appropriately. The spectra were then wavelength calibrated against a ThAr emission-lamp. The data were rescaled to a linear velocity scale appropriate for the [O~\textsc{iii}] emission, and corrected to heliocentric velocity, $v_{hel}$. The reduced spectra are presented in figures \ref{fig:1to5} and \ref{fig:6to10} as position-velocity (PV) arrays.

%\includegraphics[scale=1.3]{lowcontrast.eps}
%\caption{High resolution image of PN HaTr~4 taken on the NTT-EMMI spectrograph in \ha\ + \NII{}. The central star is visible at 16:45:00.2 -51:12:22.0}
%\label{fig:hatr4hi}

%\begin{figure*}
%\centering
%\includegraphics[scale=1.5]{imslit.eps}
%\label{fig:allslitso3}
%\caption{Diagram showing all slit positions taken on the VLT in \OIII{}. Slit width is 0.6\arcsec{}.}
%\end{figure*}

%\section{Spectral Analysis}

The PV arrays presented in figure \ref{fig:1to5} are consistent with a nebula extended in the North-South direction, as indicated by the NTT \ha{}+[N~\textsc{ii}] image shown in figure \ref{fig:allhatr4}, with slits 1, 3 and 5 showing velocity ellipses. However, some data are missing as the slit is not quite long enough to cover the full North-South extent of the nebular emission (as indicated by the sudden cut-off at the top and bottom of the PV arrays, e.g. at cross-section 13.5\arcsec{} in figure 2(e)). The PV arrays from both slits 1 and 3 (figure \ref{fig:1to5}(a) and (c)) appear to be closed at their northernmost extent (as supported by the nothern extension of slit 3, shown in slit 2, figure \ref{fig:1to5} (b)) but open in the South; however, the data from the southern extension to slit 3 - slit 4, figure \ref{fig:1to5}(d) - clearly shows that this PV array is in fact closed.% Because of the missing emission and lack of futher `extension' slits, it is difficult to confirm if the velocity ellipses shown in figures \ref{fig:1to5} (a) and (d) are truly open to the South, or are indeed closed.

The presence of two separate velocity components (corresponding to the red- and blue-shifted sides of a velocity ellipse) is consistent with front and back walls of an expanding hollow shell extended along the slit length (i.e. in a North-South direction) - this orientation is perpendicular to the East-West orientation previously indicated by the imagery presented in figure \ref{fig:allhatr4}. The emission from the central (equatorial) region of this shell is significantly brighter than that from the North and South ends, and it is this equatorial region which gives the impression of an East-West `bow-tie' shape, as seen in figure \ref{fig:allhatr4}(a).

The PV arrays presented in figure \ref{fig:6to10} are obtained from slit positions perpendicular to those in figure \ref{fig:1to5}. They each show a closed velocity ellipse, confirming the presence of a shell extending in a North-South direction. Slit 7 appears to have a larger FHWM for the blue-shifted component at cross-section 0\arcsec{}, whereas slit 9 appears to have a larger FWHM for the red-shifted component at cross-section 0\arcsec.

\section{Spatio-Kinematical Reconstruction}

%shell thickness  - is thickness consistent between model and obs? consistent with hubble flow?  A thermal width thing?
%Chosen model's good points and failures

Using the astrophysical modelling program \textsc{shape}\footnote{http://bufadora.astrosen.unam.mx/shape/} \citep{steffen06,steffen11}, a spatio-kinematical model was developed in order to reconstruct the nebular morphology of HaTr~4 based on both the high-resolution \ha{}+[N~\textsc{ii}] imagery shown in figure \ref{fig:allhatr4}(a) and the high-resolution, spatially resolved [O~\textsc{iii}] spectra shown in figures \ref{fig:1to5} and \ref{fig:6to10}.

% with the intention of acquring a best-fit inclination to the line of sight of the nebula, confirm its bipolarity in the North-South direction, and acquire an expansion velocity and kinematical age. 

\begin{figure*}
\centering
\includegraphics[]{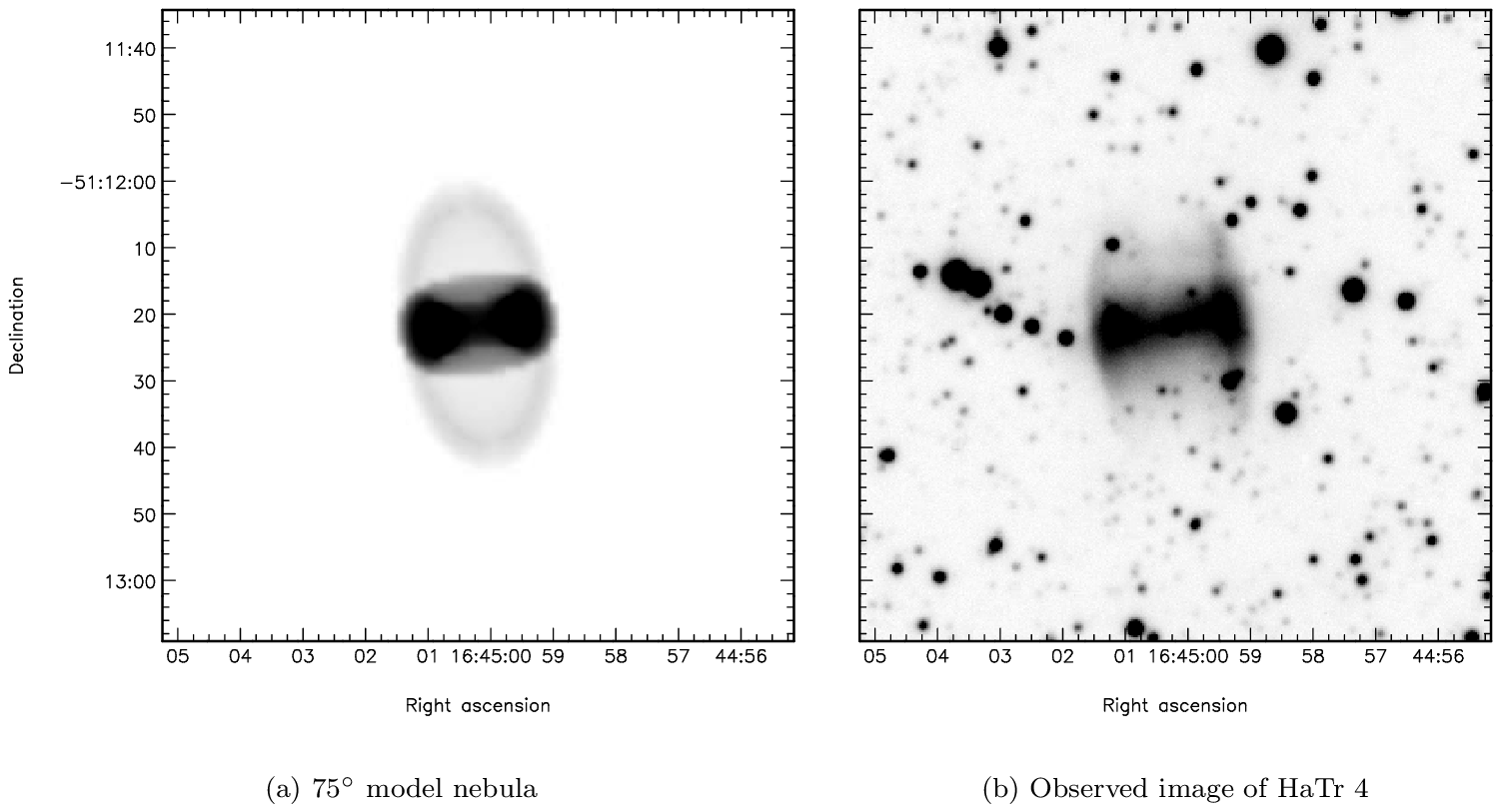}
\caption{\textsc{shape} model of HaTr~4 at the best-fitting inclination of 75\degr{}, with the real image for comparison.}
\label{fig:modelneb}
\end{figure*}

\begin{figure*}
\centering
\includegraphics[]{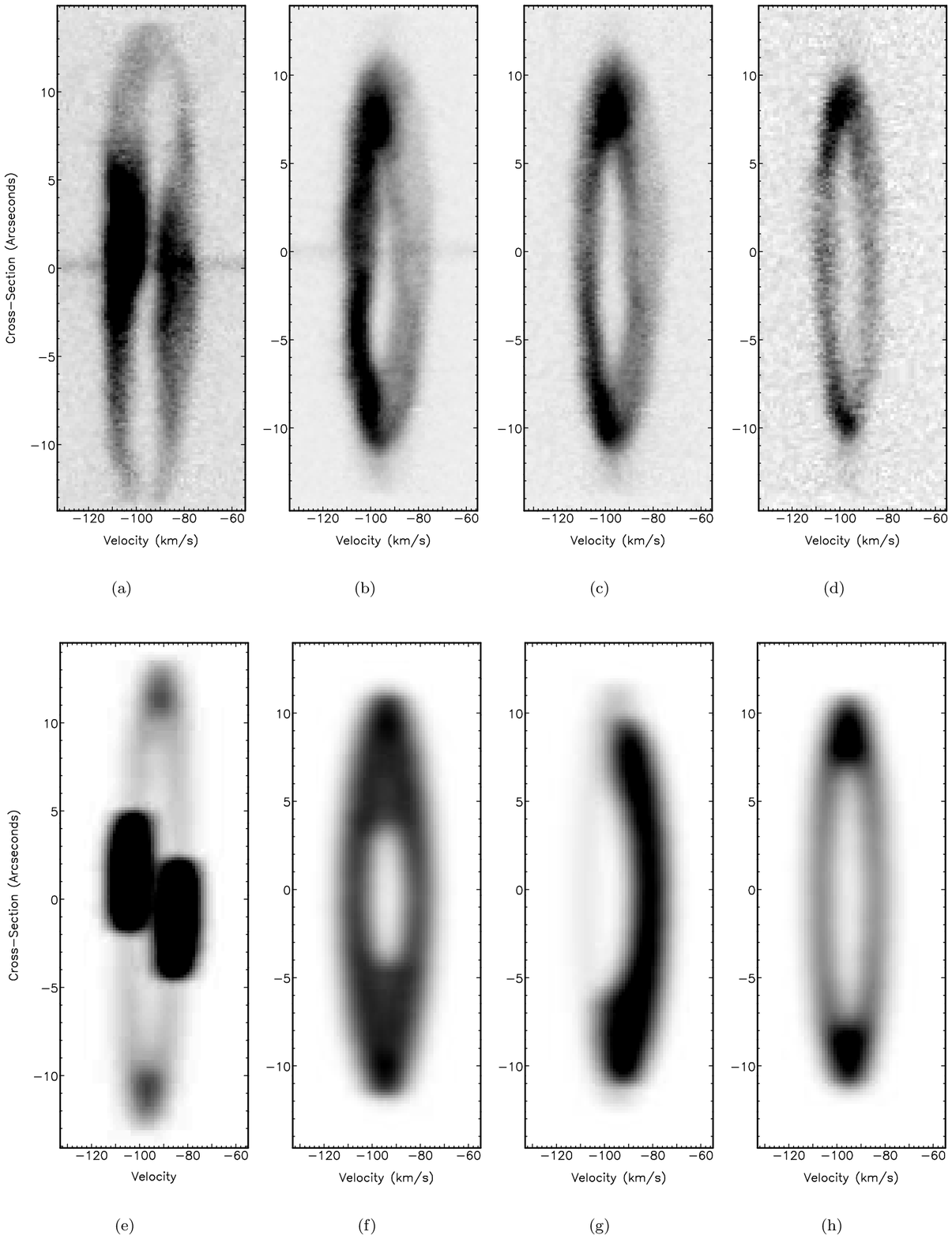}
\caption{Top row ((a)--(d)) reproduces the observed spectra for slit positions 3, 8, 9 and 10 (see figures \ref{fig:1to5} and \ref{fig:6to10}). Bottom row ((e)--(h)) shows the equivalent \textsc{shape} models, at a best fitting inclination of 75\degr{} relative to the line of sight, to show the differences between cuts through the major axis, equatorial ring, and the outer shell of the nebula. Slit positions are as shown in \ref{fig:allhatr4}(b)}
\label{fig:modelspectra}
\end{figure*}

A wide range of different morphologies and orientations for the nebular shell and waist were applied to the model of HaTr~4. The nebular expansion velocity was assumed to be a Hubble-type flow, but the scale velocity was considered a free parameter in the modelling ([r/r$_{o}$]*k, where r is radius from the central star, r$_{o}$ is the inner equatorial radius , and k is the velocity at r$_{o}$). The best-fitting basic model, determined by eye, was an elongated ovoid nebular shell encorporating a thicker equatorial ring, an image of which is shown in figure \ref{fig:modelneb} at the same scale, P.A. (7\degr{}), and inclination to the line of sight as the image shown in figure \ref{fig:allhatr4} and presented again here for comparison. The synthetic PV arrays corresponding to this best-fitting model are presented in figure \ref{fig:modelspectra}. The model accurately reproduces the two velocity components seen in all of the slit positions shown in figures \ref{fig:1to5} and \ref{fig:6to10}, corresponding to the front (blue-shifted) and back (red-shifted) walls of the main nebular shell. The model also recreates the bright emission associated with the nebular waist, appearing as the two bright velocity components around cross-section 0\arcsec{} (see figure \ref{fig:modelspectra}(e)). 
 
An ovoid model was favoured over a more cylindrical structure, as a cylindrical shell produced a nebula which was too `straight-edged' compared to the observed imagery, and did not create the roughly elliptical profile followed by the two spectral components observed in slits 1 to 10. Similarly, a tighter waist was also discounted as it gave both the nebular image and spectra a `figure-of-eight'-like appearance, with the nebular shell protruding from the central `bow-tie' rather than it being a smooth transition i.e. it is not `wasp-waisted', akin to other such close binary central star planetary nebulae (CSPNe) whose waist radius is comparable to that of its bipolar lobes such as Abell 41 \citep{jones10a}. If the shell thickness parameter was set to greater than 5\arcsec{}, it created broader components at the northern and southern tips of the model spectra than are observed in the data. A ring was applied to the equatorial region of the model in order to increase the apparent thickness of the shell at low latitudes. The `bow-tie' shape could not be accurately reproduced by simply enhancing the brightness of the shell in this region. Moreover, the increased physical thickness also leads to broader FWHM spectral components from this equatorial region due to the larger range of velocities present.

%\subsection{Constraining the nebular inclination}

Once the basic model was determined, its inclination was varied in order to assess over what range a satisfactory fit to the data could be found. Comparison between observed and model imagery constrains the inclination of the nebula to be between 65\degr{} and 80\degr{} -- below 65\degr{} the central `bow tie' becomes too extended in the North-South direction, and its centre becomes too narrow as less of the ring overlaps in the line of sight. Above 80\degr{}, the ring becomes too `tube-like' in shape rather than the apparent bow tie as the model nebula becomes close to lying in the plane of the sky. Comparison between modelled and observed spectroscopy further constrains the inclination to be 75\degr{}$\pm$5\degr{}, as other inclinations did not replicate the observed brightness offset between red- and blue-shifted velocity components seen in slits 1 to 5.

The key parameters of the best-fit model are summarised in table \ref{tab:params}. The determined expansion velocity relative to the central star of $V_{exp}$=13$\pm$2\kms{} in the equatorial plane is fairly typical for a PN \citep{weinberger89}, and the systemic velocity ($V_{sys}$=-94$\pm$4\kms{}) is in good agreement with that of \citeauthor{beaulieu99} (\citeyear{beaulieu99}; $V_{sys}$=-97$\pm$15\kms{}). The kinematical age per unit distance, as determined by the model parameters, is found to be 2740$\pm$420 years kpc$^{-1}$. Using his surface brightness-distance relationship, Frew \citeyearpar{frew08} finds a distance to HaTr~4 of roughly 3 kpc; using this value, we derive a kinematical age of 8200$\pm$1300 years for HaTr~4 and an equatorial radius of approximately 0.22 pc.

\begin{table}
\centering
\caption{Table of the main parameters of nebular ring in the best-fitting \textsc{shape} model of HaTr~4 shown in figure \ref{fig:modelneb} (errors are the range over which the parameter could be altered and still be considered a reasonable fit to the observed image and spectra).}
\begin{tabular}{ll}
\hline
\bf{Parameter} & \bf{Value}\\
\hline
Semi-major axis (to outer edge) & 12.65$\pm$0.5\arcsec{}\\
Semi-minor axis (to outer edge) & 7.5$\pm$0.5\arcsec{}\\
Shell thickness & 5$\pm$1\arcsec{}\\
Equatorial expansion velocity, $V_{exp}$ (r = 7.5\arcsec{}) & 13$\pm$2\kms{}\\
Heliocentric systemic velocity, $V_{sys}$ & -94$\pm$4\kms{}\\
Inclination, $i$ & 75$\pm$5\degr{}\\
\hline
\end{tabular}
\label{tab:params}
\end{table}

Figure \ref{fig:modelspectra}(e) shows the model spectrum corresponding to the observed PV array from central slit 3 (reproduced above it in figure \ref{fig:modelspectra}(a)) at the best-fitting inclination of 75\degr{}. The model PV array nicely confirms that the splitting of bright components shown in the spectrum is indeed due to an elongated shell possessing a thicker and brighter equatorial ring, as the relative vertical offset between the two bright components is dependent on nebular inclination. Figures \ref{fig:modelspectra}(f), (g) and (h) shows the \textsc{shape} models of slits 8, 9 and 10 (see figure \ref{fig:allhatr4}(b)) at the best-fitting inclination of 75\degr{} beneath their observed spectral counterparts to highlight how the brightness variations across the velocity ellipses changes depending on the shell/waist overlap as a result of inclination effects. The fact that the varations in the model do not exactly match the spectra highlights the asymmetry of the structure of HaTr~4 in reality. Indeed, in all of the spectra shown in figures \ref{fig:1to5} and \ref{fig:6to10}, the blue-shifted emission component is observed to be brighter than the red-shifted with no obvious explanation, as the best-fitting model does not replicate this trend; however, possible explanations are ISM interaction (which may also account for the thinner nature of that side in figure \ref{fig:6to10}(c)), or more likely that there is internal extinction blocking emission from the back wall.

%The velocity difference between the front and back walls of the nebular shell was determined by subtracting the heliocentric velocity of a point on the red-shifted side of the model spectra from an equivalent point on the blue-shifted side, giving an expansion velocity of \textbf{2$\Delta$v = 26\kms{}}. The angular size of the nebula was determined to be 23\arcsec{}. A standard PN distance of 1kpc was taken --- using this and the angular size of the nebula, the physical diameter was determined to be 1.72 \times 10$^{12}$ km. The kinematical age of HaTr~4 is therefore estimated to be of order \textbf{4200 years kpc$^{-1}${}}. Using a surface brightness-distance relationship, Frew \citeyearpar{frew08} has determined a distance to HaTr~4 of roughly 3 kpc; using the previously determined age per kpc, this gives HaTr~4 a kinematical age of order 12,580 years. The expansion velocity and age of the nebula are fairly standard for a PN of this size.

The \textsc{shape}-modelled nebular inclination of $75^{\circ}\pm5^{\circ}$ relative to the line of sight is approximately perpendicular to the binary plane, as determined in the work of Hillwig et al.\ (in preparation) and \citet{bodman12}. \citet{bodman12} state that the best-fit range for the binary falls between 75$^{\circ}$ and 80$^{\circ}$, but with an acceptable fit as low as 55$^{\circ}$. In knowing that the PN symmetry axis is expected to lie perpendicular to the orbital plane of the binary and having seen that the inclinations of the two \emph{are} in agreement, these results further imply that the binary is responsible for influencing the shaping of HaTr~4.

\section{Discussion}

The imaging, spectroscopy and subsequent spatio-kinematical modelling presented in this paper clearly show HaTr~4 to have an elongated, axisymmetric morphology with an equatorial enhancement consistent with a nebular ring. This work demonstrates for the first time that the symmetry axis of HaTr~4 lies in a roughly North-South orientation, as opposed to the East-West orientation inferred from previous imagery. The ``bow-tie'' appearance of the central region of the nebula is shown to result from a line-of-sight inclination effect associated with the enhanced nebular waist, rather than a bipolar structure as previously believed.

The spatio-kinematical modelling of HaTr~4 reveals fairly typical dynamical properties, with a kinematical age of order 8000 years and an equatorial expansion velocity of 13$\pm$2\kms{}. No evidence is found for extended emission beyond the main nebular shell or jet-like outflows associated with HaTr~4, such as those found in the binary-centred PNe ETHOS~1 \citep{miszalski11}, The Necklace \citep{corradi11}, Abell~63 \citep{mitchell07} and NGC~6337 \citep{garcia09}. Similarly, even once the effects of inclination have been taken into account, no obvious analogue for HaTr~4 can be found amongst the other binary-centred PNe which have been the subject of detailed spatio-kinematical study. While the morphology of HaTr~4 is axisymmetric it is not bilobed as those PNe mentioned above are, displaying more of an elliptical overall structure. Perhaps the closest morphologically to HaTr~4 are Abell~63 \citep{mitchell07} and Abell~41 \citep{jones10a}; however, both show marked differences to HaTr~4: Abell~63 has no ring-like equatorial feature, and also displays extended polar outflows which are absent in HaTr~4. Abell~41 does display an equatorial ring, but a far less extended version than the one seen in HaTr~4. Similarly, Abell~41 has a distinct bilobed structure not seen in HaTr~4. This lack of direct analogue highlights the morphologically disparate nature of the current sample of binary-centred PNe that have been studied in detail \citep{jones11}. The imaging study of \citet{miszalski09b} did, however, indicate that the most common morphological feature amongst the known sample is some form of equatorial ring - a feature we have shown to clearly be present in HaTr~4, and is probably associated with the ejection of a common envelope.

The inclination of the nebular symmetry axis is found to be $\sim$75\degr{}, consistent with the non-eclipsing nature of the binary central star. Further investigation into the central star system by Hillwig et al.\ (in preparation) and \citet{bodman12}, indicates that the inclination of the central binary plane is very close to the nebular inclination, as predicted by theories of binary-induced PN shaping \citep{nordhaus06}. HaTr~4 is one of only six PNe to have had this alignment between nebular symmetry axis and binary plane observationally shown, a finding entirely consistent with HaTr~4 having been shaped by its central binary star.

%This alignment between the nebular symmetry axis and binary plane is entirely consistent with HaTr~4 being shaped by its central binary star, making it one of only six PNe to have had this link observationally shown.

\section*{Acknowledgments}

AAT gratefully acknowledges the support of STFC through her studentship. We wish to thank Todd Hillwig for several highly beneficial discussions regarding the central star binarity of HaTr~4. We thank the staff at the ESO La Silla Paranal Observatory for their support. This work was co-funded under the Marie Curie Actions of the European Commission (FP7-COFUND).

\bibliographystyle{mn2e}
\bibliography{literature.bib}

\label{lastpage}

\end{document}